\documentclass[aps,pra,showpacs,notitlepage]{revtex4-1}

\usepackage{graphicx}

\begin{document}

\title{On the fundamental role of dynamics in quantum physics}

\author{Holger F. Hofmann}
\email{hofmann@hiroshima-u.ac.jp}
\affiliation{
Graduate School of Advanced Sciences of Matter, Hiroshima University,
Kagamiyama 1-3-1, Higashi Hiroshima 739-8530, Japan}

\begin{abstract}
Quantum theory expresses the observable relations between physical properties in terms of probabilities that depend on the specific context described by the ``state'' of a system. However, the laws of physics that emerge at the macroscopic level are fully deterministic. Here, it is shown that the relation between quantum statistics and deterministic dynamics can be explained in terms of ergodic averages over complex valued probabilities, where the fundamental causality of motion is expressed by an action that appears as the phase of the complex probability multiplied with the fundamental constant $\hbar$. Importantly, classical physics emerges as an approximation of this more fundamental theory of motion, indicating that the assumption of a classical reality described by differential geometry is merely an artefact of an extrapolation from the observation of macroscopic dynamics to a fictitious level of precision that does not exist within our actual experience of the world around us. It is therefore possible to completely replace the classical concepts of trajectories with the more fundamental concept of action phase probabilities as a universally valid description of the deterministic causality of motion that is observed in the physical world.
\end{abstract}


\maketitle

\section{Introduction}

Quantum theory is supposed to provide the most fundamental description of nature at the microscopic level. However, much confusion is caused by the fact that this fundamental description appears to involve probabilities and randomness. Recent developments in the field of quantum information have moved this problem back to the center of attention, revealing that the experts in the field find it hard to agree even on basic notions such as the nature and physical meaning of quantum states \cite{Pus12,Lei14,Rin15}. What we can agree on is that the statistics of measurement results obtained with quantum mechanical precision can be very strange, as evidenced by a growing number of paradoxes \cite{Bell,KS,LGI,box,Har92,Che}. All of these paradoxes are described by the standard formulation of quantum theory, and experiments confirm the measurement statistics predicted by the theory. The problem that causes the confusion is that the standard description provides only separate statistics for non-commuting physical properties, leaving important questions about the relations between these physical properties unanswered. 

When quantum theory was established, it was argued by Heisenberg and others that it was fundamentally impossible to know anything more about the correlation of non-commuting properties because of the unavoidable uncertainties in the preparation and the measurement of the system. This claim is primarily used to justify the absence of any explanations of how state preparation and measurement work. Instead, we are left with ad hoc postulates that seem to fit the facts, even if these facts violate our intuitive sense of reality or -perhaps more importantly - our intuitive sense of causality. Sadly, this state of affairs continues until today, despite a growing amount of research addressing the measurement problem itself. However, there might be a light at the end of the tunnel: recently, there have been several reports suggesting that we might be able to explore the physics beyond the uncertainty limit by paying closer attention to the statistics of measurements. In particular, it has been demonstrated that paradoxical statistics appear as negative values of probabilities observed in the post-selected statistics of weak measurements, and that a closely related statistical analysis of experimental data can be used to verify the observation of Ozawa that measurement disturbances might be more controllable than Heisenberg and Bohr suggested \cite{Lun09,Yok09,Gog11,Suz12,Erh12,Roz12,Rin14,Iin16}.

In previous work, I have analyzed the physical meaning of the complex probabilities that emerge from weak measurements and found that the complex phases represent the actions of unitary transformations that relate the physical properties to each other \cite{Hof11,Hof12a,Hof14a,Hof15}. The results of this analysis suggest that the origin of all non-classical fundamental relations between physical properties can be explained in terms of the action of transformations, which appears as a phase in the complex valued conditional probabilities that describe the non-classical statitistics of non-commuting physical properties. The more familiar form of deterministic relations known as classical physics emerges as an approximation in the coarse grained limit, where the imaginary and negative parts of the complex probability distributions are suppressed by a factor proportional to the ratio of Planck's constant and the coarse grained phase space area \cite{Hof12a}. This discovery raises important questions with regard to the nature of causality in quantum physics. Classical physics is completely deterministic, so that the position and velocity of a particle uniquely determine its path. Since it should be possible to explain classical physics as an approximation of the more fundamental quantum theory of motion, it would be important to explain how classical determinism emerges from the microscopic statistics of quantum properties. The previous analysis indicates that complex valued statistics may provide the right answer to this question, since the complex phases of the probabilities can be explained in terms of the action of deterministic motion \cite{Hof11}. 

Both the uncertainty principle and the results of weak measurements strongly suggest that the statistics of quantum states are only random because of the dynamics of the state preparation process. In the following, I will show that this conjecture is confirmed by the statistical properties of the complex probabilities observed in weak measurements, which I will hence refer to as action phase probabilities. Specifically, it will be shown that the time average of the action phase probability of a variable conditioned by an initial condition and an eigenvalue of energy is equal to the experimentally observed probability of that variable in the respective eigenstate of energy. It is therefore possible to identify all eigenstates with ergodic randomizations of the dynamics generated by the respective physical properties. Note that this is consistent with the historical derivation of quantum mechanics, where the state of an electron in hydrogen was identified with a time-averaged orbit around the nucleus. In addition, we can now see that the problem of state preparation is a problem of the means of control. The available laws of motion require that a system for which the energy is known will be randomized in time, and this applies by analogy to all other physical properties as well. 

The law of quantum ergodicity can explain the structure of Hilbert space and the associated operator algebra in terms of the causality relations described by action phase probabilities. In particular, non-commutativity actually represents imaginary correlations between dynamically related properties. These imaginary correlations indicate that three dynamically related properties cannot have a joint reality because the dynamics by which one property causes an observable physical effect will always change the other two properties in a way that is fully determined by the action phase probabilities that relate these properties to each other. 

The center piece of the following analysis is a new insight into the relation between the action of unitary transformations and the non-classical correlations between physical properties represented by non-commuting operators. This relation is introduced in Sec. \ref{sec:apps}, where it is shown that the complex phases that appear in quasi-probabilities defined by ordered products of projection operators correspond to the action phases that define unitary transformations in the Hilbert space formalism. In Sec. \ref{sec:time}, this general relation between complex phases and the action of transformations is applied to describe the time evolution of quantum statistics as a relation between initial and final conditions, so that the laws of motion can be separated from the statistics of the initial state. In Sec. \ref{sec:ergo}, it is then shown that the initial statistics of pure states are obtained by randomizing over the dynamics generated by the physical property which the state represents. Thus, the randomness of quantum states can be explained in terms of a randomization of the deterministic time evolution associated with state preparation. In Sec. \ref{sec:Et}, the randomness of quantum states is explained in terms of the energy-time uncertainty trade-off associated with the physical interactions of quantum state preparation. 

Sec. \ref{sec:apps} to Sec. \ref{sec:Et} form the first part of the paper, which shows that the action phases of complex probabilities provide a complete description of quantum dynamics that can explain the physics of quantum state preparation and measurement without additional postulates or axioms. In Sec. \ref{sec:class} to Sec. \ref{sec:concepts}, the relation between classical laws of motion and the action phases of the quantum formalism is explored. The results indicate that the complex phases of quantum theory can be derived from classical laws of motion, since the concept of action provides a unified description of causality that is equally valid in both quantum physics and its classical approximation. Non-classical correlations can therefore be explained in terms of dynamical relations between physical properties that can be formulated using the familiar language of classical physics. 

In Sec. \ref{sec:class}, it is shown that the action phases that appear in the standard Hilbert space formalism describe the approximate relation between the energy and the time it takes to move from $a$ to $b$. It is therefore possible to derive the complex phases of Hilbert space from the classical relation between energy and time associated with the dynamics of the system, where the modifications introduced by the quantum formalism correspond to the difference between the eigenvalues of energy and the weak value defined by the initial and final conditions. In Sec. \ref{sec:example}, the case of a free particle is discussed, and it is shown that the complex phases of the wavefunctions are identical to the action derived from the classical energy-time relations of particle motion. It is shown that the classical trajectories emerge when the weak value of energy is approximately equal to the eigenvalue. However, other contributions do not necessarily cancel completely in the ergodic average, which is the reason why particles can tunnel through barriers with a potential energy larger than the eigenvalue. In Sec. \ref{sec:concepts}, the relation between the fundamental concepts of classical physics and the fundamental concepts of quantum physics is discussed. It is pointed out that the conventional assumption that trajectories described by differential geometry are a fundamental concept of classical physics is not correct, which might be the most important source of confusion in quantum physics. The action provides an equivalent description of the observable changes of physical properties over time without the artificial requirement of infinite precision required for the mathematical definition of a trajectory as a line in time and space.  

The analysis developed in this paper shows that action phase probabilities can completely replace classical trajectories as a more fundamental description of causality and the associated dynamics of physical systems. As should be expected, classical trajectories only emerge as an approximation in the limit of low precision, which can now be identified with the action represented by the relations between energy and time for motion from an initial condition $a$ to a final condition $b$. It is therefore possible to arrive at a formulation of quantum physics that includes classical physics as a natural limit, so that we can finally identify and revise the faulty assumptions about reality that break down whenever the observable evolution of physical properties is sensitive to actions smaller than $\hbar$.

\section{Action phase probabilities}
\label{sec:apps}

In principle, the Hilbert space formalism of quantum mechanics should replace classical physics as a more fundamental description of the physical world around us. It is therefore a problem that the quantum states of Hilbert space appear to offer only an incomplete description of reality. When a specific property $\hat{A}$ is known, the system is described by an eigenstate from the complete basis set $\{\mid a \rangle\}$, but these eigenstates leave the values of a non-commuting property $\hat{B}$ undefined, since they are described as superpositions of the basis set $\{\mid b \rangle\}$. Classical physics seems to provide a more complete description of the physics in terms of a phase space point $(a,b)$ that assigns precise values to both $\hat{A}$ and $\hat{B}$ at the same time. 

Attempts to reconcile quantum statistics with classical phase space statistics often result in logical inconsistencies, culminating in the formulation of various paradoxes such as the violation of Bell's inequalities \cite{Bell,KS,LGI,box,Har92,Che}. Nevertheless, the operator formalism itself provides a natural phase space analog, since the projection operators that define the probabilities of the physical properties $a$ or $b$ can be combined by an operator product to define an expression that resembles a joint probability of $a$ and $b$ for any state $\mid \psi \rangle$ \cite{McCoy32,Kir33,Dir45},
\begin{eqnarray}
\label{eq:jprob}
\rho(a,b) &=& \langle \mid b \rangle \langle b \mid a \rangle \langle a \mid \rangle
\nonumber 
\\ &=&
\langle \psi \mid b \rangle \langle b \mid a \rangle \langle a \mid \psi \rangle. 
\end{eqnarray}
The problem with this phase space analogy is that the product of two non-commuting projectors is neither a projector nor even a self-adjoint operator, so that quantum state $\mid \psi \rangle$ generally assigns complex numbers to the joint probabilities of $a$ and $b$. It is therefore clearly impossible to explain quantum states as distributions over possible joint realities of $a$ and $b$. On the other hand, the negative real values obtained for joint probabilities of physical properties that cannot be observed jointly can actually explain the violation of statistical bounds observed in the various quantum paradoxes, and correspond to the non-classical correlations observed in uncertainty limited joint and sequential measurements \cite{Suz12,Hof15,Kin15}. 

Another criticism of the complex joint probability defined by Eq.(\ref{eq:jprob}) is that its motivation seems to be purely mathematical. Since joint measurements of $a$ and $b$ are impossible, standard quantum theory seems to suggest that it might be impossible to obtain any experimental evidence for complex joint probabilities. However, it turns out that this expectation is not correct. Firstly, it is possible to observe correlations between non-commuting observables in weak measurements, and it has been shown that the complex joint probability $\rho(a,b)$ can be observed experimentally by using a fairly straightforward background noise subtraction on a low resolution measurement of $a$ followed by a precise measurement of $b$ \cite{Lun11,Lun12,Sal13,Bam14}. A more detailed analysis of measurement statistics shows that the complex joint probability $\rho(a,b)$ can also be obtained at variable measurement strengths if the statistics of back-action errors is taken into account  \cite{Suz12,Kin15,Hof14b}. It has also been shown that the non-classical correlations described by the complex joint probabilities $\rho(a,b)$ appear directly in optimal quantum cloning, where the real part represents the experimentally observed correlations between the clones \cite{Hof12b}. The cloning result indicates that ideal cloning is represented by a non-positive linear map, so that cloning errors are necessary to compensate the negative probabilities that would necessarily emerge if a faithful copy of non-commuting properties was possible. A similar argument also applies to quantum teleportation, where the use of entanglement permits the simultaneous transfer of non-commuting properties \cite{Hir13}. In all of these processes, the causality relations that describe the transfer of physical properties to other systems can be expressed by complex probabilities, while the probabilities of actual measurement results remain positive as a result of the necessary environmental noise added during state preparation or measurement. 

Although the source of statistical uncertainty can always be traced to the environment, it is important to understand why the randomization of the dynamics always results in the specific statistics predicted by the Hilbert space formalism. Specifically, the effects of environmental noise should be described in terms of the dynamics of the system. In quantum mechanics, the formal expression for any dynamical transformation is given by a unitary operator $\hat{U}$. Although the mathematical form of this operator is very similar to the operators used to describe quantum statistics, the unitary transformation describes a fully reversible deterministic transformation of the physical properties. It achieves this by attaching a complex phase of $E t/\hbar$ to each energy eigenstate of energy $E$. Thus, the physics of the transformation is given by the action $S(E)=E t$ associated with stationary states $n$. Note that the action $S(E)=E t$ can also be used to define canonical transformations in classical physics, where the action is associated with a classical trajectory of energy $E$ in place of the stationary state. In this sense, the complex phases of Hilbert space vectors have a well-defined classical limit that does not depend on statistical considerations.  In the formalism of quantum mechanics, the canonical transformation given by the action $S_n=E_n t$ is expressed by a unitary transformation that generates a phase shift of $-S_n/\hbar$ in the eigenstate components $n$, 
\begin{equation}
\label{eq:unitary}
\hat{U} = \sum_n \mid n \rangle \langle n \mid \exp\left(-i \frac{1}{\hbar} S_n\right).
\end{equation}
Importantly, unitary operations can be expressed in terms of projection operators and are therefore related to measurement probabilities in a way that fundamentally contradicts the assumptions of classical physics. In classical physics, the action $S(E)$ describes a coordinate transformation between alternative descriptions of physical reality, where the state at time $t_0$ can be defined in terms of the physical properties $(a,b)$ and $n$ is a well-defined function of $(a,b)$. The conditional probability $P(n|a,b)$ would therefore be a delta function that assigns the correct value of $n$ to the initial and final conditions $(a,b)$. However, quantum mechanics modifies the statistical expression of causality that relates the measurement outcome $n$ to the initial conditions $(a,b)$ by relating it to the products of projection operators in Eq.(\ref{eq:jprob}) \cite{Hof14a}. As a result of this modification, the experimental probability $P(n)$ of a measurement outcome $n$ is obtained from the complex joint probability $\rho(a,b)$ by a complex-valued conditional probability that relates each combination of physical properties $(a,b)$ in the input state to the outcome $n$ \cite{Hof12a},
\begin{equation}
\label{eq:condprob}
P(n|a,b) = \frac{\langle b \mid n \rangle \langle n \mid a \rangle}{\langle b \mid a \rangle}.
\end{equation}
This expression of deterministic causality refers to combinations of physical properties that cannot be measured jointly, which is why the values of the conditional probabilities can be complex without causing any contradictions with experimentally accessible evidence. In terms of the statistical formalism of quantum mechanics, Eq.(\ref{eq:condprob}) relates the non-classical correlations between $a$ and $b$ in the initial state to the probabilities of an entirely different physical property $n$ observed in a separate measurement. Experimentally, this kind of relation between three non-commuting properties can be observed in weak measurements, where the conventional choice would be a preparation of $a$ and a post-selection of $b$. However, it is important to note that the relation between $a$, $b$ and $n$ represented by $P(n|a,b)$ is actually independent of the temporal order and describes a universal and state-independent relation between the physical properties $a$, $b$ and $n$. This can be demonstrated experimentally by preparing $b$, performing a weak measurement of $a$, followed by a final measurement of $n$. Such a measurement determines the complex joint probability of $a$ and $n$ in the state defined by $b$, which can then be converted into the conditional probability in Eq.(\ref{eq:condprob}) by the usual procedure of dividing by the marginal probability of $a$. In this measurement sequence, the preparation of $b$ followed by a weak measurement of $a$ is used to approximate the joint preparation of $(a,b)$, which is prevented by the uncertainty limits of the dynamics by which the properties $a$ and $b$ can be controlled. In this manner, weak measurements can provide insights into the physics of uncertainties that are usually hidden by the environmental noise in the interaction dynamics by which quantum systems are controlled.

Independent of experimental procedures, it is important to realize that $P(n|a,b)$ describes the fundamental causality relation between $a$, $b$ and $n$ in the Hilbert space formalism, where only the sign of the imaginary part depends on the order in which the three properties appear. In particular, $P(n|a,b)$ serves to establish the relation between measurement outcomes $n$ and input states described by $\rho(a,b)$ for all possible input states, where the physical properties $(a,b)$ serve as a parameterization of the statistics in terms of the physical properties corresponding to $a$ and $b$. The expression $P(n|a,b)$ thus applies consistently to all possible quantum states, representing a state independent description of the relation between three physical properties that replaces the deterministic dependence of $n$ on $(a,b)$ used in classical physics. It is therefore reasonable to consider $P(n|a,b)$ as a fundamental description of causality, and not just as a statistical result obtained in a specific measurement setup. In the following, this fundamental expression of relations between physical properties is used to analyze the universal patterns of quantum dynamics observed in conventional unitary transformations, as described by Eq.(\ref{eq:unitary}). The results show that the expressions $P(n|a,b)$ provide a detailed microscopic description of dynamics that applies to all quantum states and can actually explain the experimentally observed probability distributions associated with eigenstates of a given physical property in terms of a dynamical average over the causality relations that govern the dynamics of unitary transformations generated by that physical property.

Comparison of Eq.(\ref{eq:condprob}) with Eq.(\ref{eq:unitary}) shows that the relation $P(n|a,b)$ can be used to express the effects of a unitary transformation generated by $n$ on the probability of finding $b$ after preparing $a$ \cite{Hof11,Hof12a,Hof14a}, 
{\samepage
\begin{eqnarray}
\label{eq:trans}
P_{\mathrm{exp.}}(b|U(a)) &=& |\langle b \mid \hat{U} \mid a \rangle|^2
\nonumber \\
&=& P_{\mathrm{exp.}}(b|a)
\left| \sum_n P(n|a,b) \exp\left(-i \frac{1}{\hbar} S_n\right) \right|^2.
\end{eqnarray}}
It should be noted that Eq.(\ref{eq:trans}) is a highly non-trivial relation between dynamics and statistics that has no analog in classical physics. The relation shows that the complex phases of the conditional probabilities $P(n|a,b)$ have a well defined physical meaning with regard to the dynamical effects of an action $S_n$ on the statistics of $b$ in $a$. As mentioned above, the complex conditional probabilities $P(n|a,b)$ originally describe the linear relation between the probability $P(n)$ and the complex joint probability $\rho(a,b)$ of an arbitrary quantum state. There is no statistical motivation for associating the probability $P(b|U(a))$ with an absolute square of a complex conditional probability, especially since this square results in two independent sums over different values of $n$. Instead, the physics expressed by Eq.(\ref{eq:trans}) cn only be understood in terms of the action $S_n = E_n t$ of the unitary transformation given in Eq.(\ref{eq:unitary}). Specifically, the complex phases of the probabilities $P(n|a,b)$ define the action $S_n$ that maximizes the probability of obtaining $b$ \cite{Hof11},
\begin{equation}
\label{eq:action}
S(b,a,E_n) = \hbar \mbox{Arg}\left(P(n|a,b)\right).
\end{equation}
Here, $S(b,a,E_n)$ is the action of a transformation, which is known in classical physics as the reduced action of Hamilton-Jacobi theory. Sepcifically, this action describes the time needed to propagate from $a$ to $b$ along a trajectory of energy $E_n$ as its derivative in energy, $\partial S(b,a,E)/ \partial E$. It is then possible to understand how and why quantum physics modifies the statistics of physical properties: the complex conditional probabilities $P(n|a,b)$ are only zero if there exists no dynamical connection that describes a transformation from $a$ to $b$ generated by $n$. If it is possible to transform $a$ into $b$ by applying an action of $S(b,a,E_n)$, this action defines the complex phase of the conditional probability $P(n|a,b)$. It is therefore possible to derive the action phases $S(b,a,E_n)$ from classical dynamics. Specifically, a dynamical connection between $a$ and $b$ in $n$ means that the transformation distance between $a$ and $b$ can be given by the time $t$ that it takes until $S(b,a,E_n)=E_n t$. In this sense, the complex phase of the action phase probability $P(n|a,b)$ describes the transformation distance between $a$ and $b$ along the ``trajectory'' given by $n$. The classical result of $n$ as a function of $(a,b)$ is recovered approximately when the transformation distance is zero, which corresponds to a minimum of the action in energy, $\partial S/\partial E = 0$. In practical terms, classical determinism emerges after coarse graining by $\Delta E$, which reduces the conditional probabilities to zero when the transformation distance exceeds $\Delta t=\hbar /\Delta E$, leaving only real and positive conditional probabilities near the classical values of $n$ \cite{Hof12a}. In the more precise limit of resolutions beyond $\hbar$, the action describing the transformation distance between $a$ and $b$ in $n$ appears as a complex phase in the non-classical probabilities that describe the fundamental correlations between non-commuting physical properties.

Since the physics of complex probabilities in quantum mechanics is determined by the action of transformations between the physical properties, I will refer to such complex-valued probabilities as action phase probabilities. Significantly, the fundamental constant of quantum mechanics $\hbar$ is the action-phase ratio that provides the precise quantitative relation between statistical effects and dynamics that characterize all quantum effects. Quantum correlations can then be described in terms of action phase probabilities, where non-positive probabilities are a well-explained consequence of the dynamical structure, similar to the way that relativistic effects are a well-explained consequence of the non-Euclidean space-time geometry. Action phase probabilities could therefore achieve for quantum physics what the non-Euclidean space time geometry has achieved for the theory of relativity. 

\section{Dynamics and time dependence}
\label{sec:time}

Since dynamics plays such a central role in quantum statistics, it may be a problem that the standard formulation is based on ``states''. It is easy to forget that the concept of a ``state'' is meaningless unless it is explained in terms of the physical properties and the processes by which the physical properties are controlled. Above all, it is not enough to associate quantum states with specific experimental situations, since this is a very limited heuristic approach that tends to ignore the physics involved in the experimental implementation of the actual state preparation process. Specifically, it is necessary to understand the physical means by which the initial conditions of a quantum experiment are controlled. The uncertainty limits of quantum states must therefore be explained in terms of the dynamics of control, not just in terms of the mathematics by which states are described. 

On closer inspection, the origin of randomness in quantum state preparation can always be traced to very specific dynamical interactions that randomize the initial conditions of the system. In a ground state, these interactions result in cooling, while necessarily disturbing the state of motion in a way that amounts to a time average of the dynamics, also known as an ergodic average. In quantum mechanics, this loss of control over the time dependence of physical properties seems to be fundamental. The disturbance caused by a precise projective measurement of an observable also results in a randomization of the dynamics generated by the observable in question. The randomness of pure quantum states can therefore be traced back to a randomness in the time evolution of all physical properties that do not commute with the known property of the system. 

For energy eigenstates, this means that the physical properties of the system continue to evolve in time even if the overall probability distributions are all stationary. It is possible to analyze this motion in an energy eigenstate by taking a look at the two-time correlations of the physical properties. In the operator formalism, such a correlation between a physical property $\hat{A}(t_0)$ and a physical property $\hat{B}(t_0+t)$ can be expressed by an ordered product of operators, where the usual convention places operators at early times right and operators at later times left. For an energy eigenstate $\mid n \rangle$ with an energy of $E_n$, the dependence of this correlation on the time difference $t$ is given by
\begin{equation}
\label{eq:tcorr}
\langle \hat{B}(t_0+t) \hat{A}(t_0) \rangle = \langle n \mid \hat{B} \; \hat{U}(t) \hat{A} \mid n \rangle \exp\left(i \frac{E_n}{\hbar} t\right).
\end{equation}
If the unitary operation $\hat{U}(t)=\exp(-i \hat{H} t/\hbar)$ generated by the Hamiltonian $\hat{H}$ does not commute with either $\hat{A}$ or $\hat{B}$, this equation describes a non-trivial time dependence of the correlation even when the system is in an eigenstate of energy. 

We can now formulate a microscopic description of the time dependence described by the correlation in Eq.(\ref{eq:tcorr}) in terms of the action phase probabilities that relate the eigenvalues of the operators to each other. This means that we can assign an initial condition $a$ at time $t_0$ and a final condition $b(t)$ at time $t_0+t$ to describe the motion within the state of constant energy $n$. The relation between $b(t)$ and the initial condition $a$ for a trajectory $n$ with energy $E_n$ is then given by 
\begin{equation}
\label{eq:motion}
P(b(t)|a,n) = \langle b \mid \hat{U}(t) \mid a \rangle \exp\left(i \frac{E_n}{\hbar} t\right) \frac{\langle n \mid b \rangle}{\langle n \mid a \rangle}.
\end{equation}
Note that the word ``trajectory'' for $n$ is used merely to indicate that $n$ represents constants of the motion. It should not be taken to refer to a geometric description of motion in space and time, since the Hilbert space formalism of quantum mechanics replaces this kind of description by the action phase probability $P(b(t)|a,n)$. 

In classical physics, $(a,n)$ would provide a complete set of initial conditions for the motion, with a well-defined value of $b(t)$ for any specific propagation time $t$, where the time $t$ needed to get from $a$ to $b$ can be expressed as the energy gradient of the action $S(b,a,E_n)$. Action phase probabilities describe the dynamics of $b(t)$ as a time-dependent change in the transformation distance described by the action phases. This time evolution can be illustrated by the action phase difference between two different trajectories $n$ and $m$,
\begin{eqnarray}
\label{eq:prop}
S(b(t),a,n) &-& S(b(t),a,m) = (E_n-E_m)t + \hbar \mbox{Arg}\left(\frac{\langle b \mid m \rangle \langle m \mid a \rangle}{\langle b \mid n \rangle \langle n \mid a \rangle}\right)
\nonumber \\[0.2cm]
&=& (E_n-E_m)t + S(b(t_0),a,n) - S(b(t_0),a,m).
\end{eqnarray}
As explained above, the classical limit emerges by coarse graining, where the complex probabilities cancel except when the time of propagation is equal to the energy derivative of the action, $t=\partial S(b(t_0),a,E)/\partial E$, which happens to be one of the results of classical Hamilton-Jacobi theory. 

In classical physics, the time dependence of $P(b(t)|a,n)$ would consist of a sudden rise in probability at the arrival time $t(b,a,E)$ and a sudden drop immediately after. Action phase probabilities represent a very different form of dynamics, characterized by the energy of the system. For a given value of $b$, the time dependence can be described by a differential equation that is very similar to the time-dependent Schr\"odinger equation,
\begin{equation}
\frac{\partial}{\partial t} P(b(t)|a,n) = \frac{i}{\hbar}\left(E_n - H(b,a,t)\right) P(b(t)|a,n),
\end{equation}
where $H(b,a,t)$ is the weak value of energy for the initial condition $a$ and the final condition $b(t)$. In terms of the Hamiltonian $\hat{H}$,
\begin{equation}
\label{eq:weakH}
H(b,a,t) = \frac{\langle b \mid \hat{U}(t)\; \hat{H} \; \mid a \rangle}{\langle b \mid \hat{U}(t)\mid a \rangle}.
\end{equation}
In particular, this relation indicates that the time derivative of the action phase is given by the difference between the energy eigenvalue $E_n$ and the real part of the time dependent weak value of energy,
\begin{equation}
\label{eq:timeS}
\frac{\partial}{\partial t} S(b(t),a,n) = E_n - \mbox{Re}\left(H(b,a,t)\right).
\end{equation}
We can compare this result to the classical expectation of a specific energy dependent propagation time $t(b,a,E)$ that the system needs to get from $a$ to $b$ at an energy of $E$. The weak value of energy can be understood as an estimate of the energy needed to get from $a$ to $b$ in time $t$, so Eq.(\ref{eq:timeS}) relates to the inverse problem of assigning an energy to a specific propagation time. Specifically, the time evolution of the action phase is stationary whenever the energy $E_n$ matches the energy needed to get from $a$ to $b$ within the alloted time $t$. At earlier times and at later times, there will be a mismatch between $E_n$ and the weak value $H(b,a,t)$, and this mismatch results in temporal oscillations of the action phase. 

The classical limit of action phases is obtained by coarse graining. In the present case, it is possible to coarse grain the time resolution. For a time uncertainty of $\Delta t$, the probability $P(b(t)|a,n)$ vanishes if the difference between $E_n$ and the real part of the weak value $H(b,a,t)$ is much greater than $\Delta E = \hbar/\Delta t$. The result is indistinguishable from the classical time dependence if the weak value $H(b,a,t)$ changes by more than $\Delta E$ during the time $\Delta t$, leaving only an approximately Gaussian peak of width $2 \Delta t$ around the expected arrival time obtained from $E_n=H(b,a,t)$.

The discussion above shows that action phase probabilities provide an alternative description of the microscopic laws of motion that is fundamentally different from the geometric trajectories of classical physics. In the coarse grained limit, geometric trajectories appear as a good approximation, but at the microscopic level, action phase probabilities describe a completely different kind of causality. The main difference is found in the relation between time and energy, which is directly addressed by the simultaneous assignment of a propagation time $t$ and an energy $E_n$ in the action phase probability $P(b(t)|a,n)$. In classical physics, this assignment is redundant because the energy is a function of propagation time, $E=H(b,a,t)$, and the propagation time is a function of energy, $t=t(b,a,E)$. In quantum mechanics, it turns out that these are approximations that are only valid in the limit of actions larger than $\hbar$, as defined by the energy-time uncertainty product, $\Delta E \Delta t \gg \hbar$. 

In the Hilbert space formalism, the peculiar relation between time and energy is expressed in terms of the Hamilton operator $\hat{H}$, which is simultaneously an observable and a generator of the dynamics. In the time dependent action phase probabilities, this dual role is separable, resulting in three distinct parts,
\begin{equation}
\label{eq:parts}
P(b(t)|a,n) = \underbrace{\langle b \mid \hat{U}(t) \mid a \rangle}_{t \;\; \mbox{only}} \exp\left(i \frac{E_n}{\hbar} t \right) \underbrace{\frac{\langle n \mid b \rangle}{\langle n \mid a \rangle}}_{E_n \;\; \mbox{only}}.
\end{equation}
This separation of time dependence and energy suggests a much greater similarity between the dynamics observed at different energies than the classical formalism suggests. Specifically, we can expect to find signatures of all energies in the action phase probability conditioned by only a single energy $n$. The energy dependence of the time evolution of $b$ emerges only as a result of the action phase factor $E_n t$ that links the time dependence of $\hat{U}(t)$ with the energy dependence of $\langle n \mid b \rangle$. 

\section{Ergodic averaging}
\label{sec:ergo}

In principle, action phase probabilities can be observed in weak measurements, e.g. by preparing $a$ at time $t_0$, performing a weak measurement of the energy eigenstate $n$ at any time in between, and finally measuring $b$ at time $t_0+t$. The weak measurement is necessary because a strong projective measurement of energy would completely eliminate the time dependent information about $a$. The mechanism by which this happens is the measurement interaction, which results in a randomization of the dynamics generated by the target observable $\hat{H}$. If the measurement result $n$ is discarded, the measurement process can be represented by a randomization of the dynamics, resulting in the dephased density matrix $\hat{\rho}_{\mathrm{out}}$,
\begin{eqnarray}
\hat{\rho}_{\mathrm{out}} &=& \lim_{T \to \infty} \frac{1}{T} \int_0^T \hat{U}(t) \mid a \rangle \langle a \mid \hat{U}^\dagger(t) dt
\nonumber \\
&=& \sum_n |\langle n \mid a \rangle|^2 \mid n \rangle \langle n \mid.
\end{eqnarray}
A precise measurement of $n$ requires a complete randomization of the dynamics generated by the associated observables. The statistics of an eigenstate of $n$ observed after the measurement thus represents an ergodic average over the dynamics described by $\hat{U}(t)$. Since the action phase probabilities describe the time dependence of the contribution of $n$ in the state $a$ explicitly, it is possible to find the ergodic probabilities $P_{\mathrm{erg.}}(b|n)$ by averaging over the time $t$ in the fundamental relation between $a$, $n$ and $b(t)$,
\begin{equation}
\label{eq:ergo}
P_{\mathrm{erg.}}(b|n) = \lim_{T \to \infty} \frac{1}{T} \int_0^T P(b(t)|a,n) dt.
\end{equation}
The probability distribution of $b$ in an energy eigenstate $n$ can therefore be explained as the result of ergodic averaging caused by a randomization of the dynamics along the trajectory defined by $n$ during the state preparation process. 

As mentioned above, the difference between action phase probabilities and the classical approximation based on geometric trajectories is that action phase probabilities can be separated into a time dependent and an energy dependent contribution linked only by an action phase factor of $E_n t$. It is therefore possible to express the time dependent part of the action phase probability by a ratio of the action phase probability at $t$ and the corresponding action phase probability at $t=0$. Using $b=b(t=0)$ to indicate the instantaneous relation between $a$, $b$ and $n$,
\begin{equation}
\frac{P(b(t)|a,n)}{P(b|a,n)} = \frac{\langle b \mid \hat{U}(t) \mid a \rangle}{\langle b \mid a \rangle} \exp\left(i \frac{E_n}{\hbar} t\right).
\end{equation}
The ergodic average of this expression corresponds to a Fourier transform of the unitary operator $\hat{U}(t)$, which results in a projector onto the eigenstate (or eigenspace) with energy $E_n$, 
\begin{equation}
\lim_{T \to \infty} \frac{1}{T} \int_0^T \frac{P(b(t)|a,n)}{P(b|a,n)} dt = \frac{\langle b \mid n \rangle \langle n \mid a \rangle}{\langle b \mid a \rangle}.
\end{equation}
By definition, this result is equal to the action phase probability $P(n|a,b)$, which means that the ergodic average of the time dependent action phase probability $P(b(t)|a,n)$ can be expressed as a product of two instantaneous (and hence time independent) action phase probabilities,
\begin{equation}
\label{eq:apperg}
\lim_{T \to \infty} \frac{1}{T} \int_0^T P(b(t)|a,n) dt = P(n|a,b) P(b|a,n).
\end{equation}
Importantly, this result applies equally for all initial conditions $a$, since the dependence on $a$ is eliminated in the ergodic average. The law of quantum ergodicity thus states that the ergodic average represented by a stationary state $n$ defines the probability of $b$ as the product of two action phase probabilities that express the universal relation between $a$, $b$ and $n$ \cite{Hof14a},
\begin{equation}
\label{eq:querg}
P_{\mbox{erg.}}(b|n) = P(n|a,b) P(b|a,n),
\end{equation}
where $a$ can be any condition that is consistent with both $n$ and $b$.  

Importantly, the ergodic probability $P_{\mbox{erg.}}(b|n)$ represents the standard measurement probabilities of quantum mechanics, usually given in their Hilbert space form as
\begin{equation}
P_{\mbox{erg.}}(b|n) = |\langle b \mid n \rangle|^2.
\end{equation}
The analysis given above shows that this definition of probabilities in Hilbert space can be explained in terms of the randomization of the dynamics along the trajectory defined by $n$ that necessarily occurs during the preparation of a state with the property $n$. This randomization can be factorized in such a way that the ergodic probabilities are given by products of time-independent complex probabilities, which corresponds to Born's rule in the Hilbert space formalism. In fact, it is possible to derive the complete Hilbert space formalism from the law of quantum ergodicity as given by Eq.(\ref{eq:querg}) and a second condition that ensures the reversibility of transformations \cite{Hof12a,Hof14a}. Thus the Hilbert space formalism is simply a less intuitive way of encoding the causality relations described by action phase probabilities.

\section{Energy-time uncertainty of quantum state preparation}
\label{sec:Et}

In the previous section it was shown that eigenstates of a physical property describe dynamically randomized states corresponding to an equal distribution of time along the trajectory described by the corresponding property. This result explains the dynamics of quantum state preparation: in order to precisely control a physical property $n$, it is necessary to interact with the physical system in such a way that the dynamics generated by the property $n$ is completely randomized. This rule is not just a technical restriction of control, but represents an aspect of the fundamental laws of causality described by action phase probabilities. 

In the classical limit, physical systems are never controlled with a precision that even remotely approaches the limit set by $\hbar$, and that is the reason why we have been ignorant of the correct form of causality for so long. In fact, the standard formulation of quantum theory is somewhat unrealistic, since it implicitly suggests that the typical preparation of a quantum state involves the precise control of a physical property $n$. A more realistic explanation of the process of quantum state preparation would acknowledge that most systems are controlled by a sequence of interactions which results in partial control of all degrees of freedom. This situation can be described by partial ergodic randomization: starting from the initial situation $a$, the initial energy uncertainty can be reduced by randomizing the dynamics over a finite time distribution with a time uncertainty of $\Delta t$. 

To understand the dynamics of partial ergodic randomizations, it may be good to recall that the initial probability distribution of $b(0)$ in $a$ is given by a marginal of the action phase probability,
\begin{equation}
P_{\mathrm{exp.}}(b|a) = \sum_n P(b(0)|a,n) P_{\mathrm{exp.}}(n|a).
\end{equation}
A finite resolution measurement of $n$ will cause a finite randomization of the dynamics and result in a Bayesian update of $P(n|a)$. Quantum ergodicity explains the relation between these two processes in terms of the probability density $G(t)$ that describes the precise form of the dynamical randomization. Specifically, the normalized distribution function $G(t)$ replaces the limit to infinity for $1/T$, so that the action phase probability after the partial ergodic randomization reads
\begin{equation}
P_{\Delta t} (b(t)|a,n) = \int G(t^\prime) P(b(t-t^\prime)|a,n) dt^\prime.
\end{equation}
This action phase probability is still complex and time dependent, since the energy resolution $\Delta E$ associated with a partial ergodic randomization is necessarily finite. The precise form of an optimized energy measurement with a back-action given by $G(t)$ is obtained by considering the amount of dephasing induced between different energy eigenstates by the partial ergodic randomization,
\begin{equation}
D(n,m) = \int G(t) \exp\left(-i \frac{(E_n-E_m)}{\hbar} t\right) dt.   
\end{equation}
In an optimal measurement of energy, the final state will be a pure state with an energy spectrum centered around the measurement outcome $E_{\mathrm{prep.}}$. To construct this state, it is necessary to combine the dynamical randomization with a Bayesian update of the energy distribution. Such an update can be obtained from the original state $a$ by multiplying the amplitudes of all energy components with a decoherence factor of $D(n,\mbox{prep.})$ and normalizing the result. In the statistics of $n$, this corresponds to a conventional statistical update with likelihoods proportional to $|D(n,\mbox{prep.})|^2$,
\begin{equation}
P_{\Delta t} (n|a) = \frac{|D(n,\mbox{prep.})|^2 P(n|a)}{\sum_m |D(m,\mbox{prep.})|^2 P(m|a)}.
\end{equation}
Since $D(n,m)$ is the Fourier transform of the temporal distribution $G(t)$, the energy resolution $\Delta E$ and the temporal uncertainty $\Delta t$ of the partial ergodic randomization satisfy an uncertainty relation given by
\begin{equation}
\Delta E \Delta t \geq \frac{\hbar}{\sqrt{2}}.
\end{equation}
Note that the factor of $\sqrt{2}$ distinguishes this uncertainty relation from other uncertainty relations because the energy resolution is determined by the squared Fourier transform of the probability distribution of time $t$. This asymmetry between the statistics of the generator $\hat{H}$ and the statistics of the dynamical parameter $t$ is a direct result of the explanation of quantum coherence in terms of the dynamical randomness represented by action phase probabilities.

The quantum state $\psi(a,\Delta t)$ obtained by the partial ergodic randomization of $a$ can be used as a new initial state characterized by a complex valued joint probability of
\begin{eqnarray}
\rho(b,n) &=& P(b|\psi(a,\Delta t), n) P(n|\psi(a,\Delta t))
\nonumber \\
&=& P_{\Delta t} (b|a,n) P_{\Delta t} (n|a).
\end{eqnarray}
Importantly, this new state combines information about the conserved quantities $n$ with information about the dynamical variables and therefore corresponds more closely to the kind of situation described by classical physics. The proper relation between the fundamental causality relations represented by action phase probabilities and the approximate causalities described by classical equations of motion should therefore be understood in terms of partial ergodic randomizations, where a sequence of observations results in an approximate estimate of both dynamical variables and conserved quantities, while the interactions associated with the observations cause a dynamical randomization that usually results in effective energy-time uncertainty products much larger than $\hbar$. We should keep in mind that our actual experience of classical trajectories is limited to very crude estimates of the physical properties that necessarily neglect a large part of the actual interaction dynamics by which these estimates are obtained. A microscopic description by action phase probabilities is therefore completely consistent with the macroscopic observation of motion that is usually associated with classical physics.

\section{Action in the classical limit}
\label{sec:class}

The classical description of causality by differential geometry breaks down in the limit of small actions. However, the effects of small actions are only observed at very high resolution, and this requires a control of the physical system that is very difficult to implement. In most cases, it is possible to coarse grain the action phase probabilities, leaving only the contributions close to the point where the differentials of the action phase is zero in time and in energy. 

Interestingly, the classical limit only applies in situations where the experimentally observable probabilities vary much more slowly than the action phases, so that the time and energy dependence of the action phase can be separated from the time and energy dependence of the total probability. As a result, it appears as if causality and statistics are separate in the classical limit, despite the fact that the action phase probabilities express a fundamental relation between the two. 

The actions that describe the complex phases of action phase probabilities can be separated according to Eq.(\ref{eq:parts}). The first action is a function of time and is given by 
\begin{equation}
S_t(b,a,t) = \hbar \mbox{Arg} \left(\langle b \mid \hat{U}(t) \mid a \rangle\right).
\end{equation}
This action corresponds directly to Hamilton's principal function, where $a$ describes the initial conditions to which this specific solution of the Hamilton-Jacobi equation applies. In quantum mechanics, the time derivative of this action is equal to the negative real part of the weak value of the Hamiltonian given by Eq.(\ref{eq:weakH}),
\begin{equation}
\frac{\partial}{\partial t} S_t(b,a,t) = - \mbox{Re} \left( H(b,a,t) \right).
\end{equation}
In the classical Hamilton-Jacobi equation, the Hamiltonian is expressed as a function of position and momentum, where the momentum is defined as the spatial derivative of the principal function. However, this representation of the Hamiltonian is merely a possible method of calculating the classical energy for a system moving from $a$ to $b$ within a time of $t$. In the classical limit, the weak value $H(b,a,t)$ is therefore identified with the observable value of the energy given by the eigenvalue $E_n$. This identification is indeed obtained from the condition that the time derivative of the total phase is zero, as given in Eq.(\ref{eq:timeS}). Therefore, the classical approximation describing the arrival time $t$ at which the action phase is stationary is defined by the equality of weak value and eigenvalue,
\begin{equation}
\label{eq:classE}
\mbox{Re} \left(H(b,a,t)\right) \approx E_n.
\end{equation}
The time dependence of $S_t(b,a,t)$ thus provides an estimate of energy that is approximately equal to the actual energy in the classical limit. 

The other contributions to the action in Eq.(\ref{eq:parts}) are given by $E_n t$ and by an energy dependent action that can be obtained from the phase coherence of the energy eigenstates $n$ with
\begin{equation}
S_E(b,a,E_n) = \hbar \mbox{Arg} \left(\langle b \mid n \rangle \langle n \mid a \rangle\right).
\end{equation}
The total action can then be written as a sum of the three contributions,
\begin{equation}
\label{eq:3S}
S(b(t),a,n) = S_t(b,a,t) + E_n t - S_E(b,a,n).
\end{equation}
If the absolute values of the action phase probabilities vary more slowly than the action phases themselves, the approximate solution of the ergodic integral in Eq.(\ref{eq:ergo}) can be obtained by integrating the phase dependence in the vicinity of the classical solution given by Eq.(\ref{eq:classE}). Since the integral results in real and positive probabilities, the action phase has to be approximately zero in the vicinity of the classical result. Therefore the classical relation between the time and energy dependent parts of the action is given by
\begin{equation}
\label{eq:W}
S_E(b,a,n) \approx S_t(b,a,t) + E_n t.
\end{equation}
This is precisely the relation between Hamilton's principal function and the reduced action. Thus the complete Hamilton-Jacobi formalism can be derived from the approximation of quantum ergodicity obtained when coarse graining removes all contributions with action phases that are not stationary in time. 

In the classical limit, the energy is continuous and the relation between $S_t$ and $S_E$ given by Eq.(\ref{eq:W}) is a Legendre transformation between the continuous functions of $t$ and of $E$, respectively. Therefore, it is possible to derive the time of propagation from $a$ to $b$ along a trajectory of energy $E$ from the energy derivative of $S_E$,
\begin{equation}
\frac{\partial}{\partial E} S_E(b,a,E) = t(b,a,E).
\end{equation}
In the quantum mechanical case, the derivative in energy should be replaced by an actual difference between energy levels, resulting in an equivalent estimate of propagation time given by 
\begin{equation}
t_{nm} = \frac{S_E(b,a,E_n) - S_E(b,a,E_m)}{E_n-E_m}.
\end{equation}
The time $t_{nm}$ provides an estimate of the time $t$ at which the energies $E_n$ and $E_m$ contribute most to the probability of $b(t)$, as discussed in section \ref{sec:time}. In the classical limit, this estimate can be identified with the actual time of propagation $t$, 
\begin{equation}
t_{nm} \approx t.
\end{equation}
Thus, the classical action formalism can be derived by identifying the conditions where the action phases vary only slowly in both time and energy. Although the fundamental laws of motion expressed by action phase probabilities require that time and energy have no joint reality, the action defines approximate relations that are valid at scales of energy and time that are much cruder than $\hbar$. In that limit, it becomes possible to approximate causality by differential equations, thus separating the dynamics from the interactions by which physical properties are observed.

\section{Laws of motion for a single particle}
\label{sec:example}

It may be good to provide a simple practical illustration of the application of action phase probabilities to the relation between dynamics and conserved quantities. Clearly, the most accessible case is that of a non-relativistic particle in free space, where there are no unsolved mathematical problems to complicate the discussion. This simple example may thus clarify how action phase probabilities change the fundamental concepts of motion and causality. 

Since the action phase probability $P(x(t)|x_0,p)$ that describes the motion of a particle of mass $m$ in free space is normalized in $x$, it is easy to find its correct form even if $\langle x \mid \hat{U}(t) \mid x_0 \rangle$ is not available from other sources. The explicit form for an initial momentum of $p$ reads
\begin{equation}
\label{eq:freeApp}
P(x(t)|x_0,p) = \sqrt{\frac{m}{2 \pi \hbar t}} \exp\left(i \frac{m}{2 \hbar t} (x-x_0-\frac{p}{m} t)^2 - i \frac{\pi}{4}\right).
\end{equation}
For the purpose of describing the dynamics, the action phase of this probability can be separated according to Eq.(\ref{eq:3S}), where the energy eigenvalue is given by
\begin{equation}
E_p = \frac{p^2}{2 m}.
\end{equation}
The time independent part of the action phase is proportional to the product of $p$ and $(x-x_0)$, which originates from the phase difference between $x$ and $x_0$ in the eigenstate of $p$ and corresponds to the reduced action of a free particle. Using $E_p$, the energy dependent action is
\begin{equation} 
S_E(x,x_0,E_p) = \sqrt{2m E_p}\; (x-x_0). 
\end{equation}
Based on this result and on Eq.(\ref{eq:3S}), it is possible to identify the time dependent action in Eq.(\ref{eq:freeApp}). The result reads 
\begin{equation} 
S_t(x,x_0,t) = \frac{m (x-x_0)^2}{2 t} - \frac{\hbar \pi}{4}. 
\end{equation}
Note that this expression differs from Hamilton's principal function by an offset of $\hbar \pi/4$ associated with the normalization of the complex probability. 

We can now confirm that the spatial distribution of free particles is given by an ergodic average of the dynamics described by Eq.(\ref{eq:freeApp}). Importantly, the main contribution to the integral originates from the vicinity of the classical propagation time given by $t_c=p/(m(x-x_0))$. Therefore it is possible to calculate the ergodic averages for all final positions between $x_0$ and $p T/m$ by a partial ergodic integral of length $T$ around the classical time $t_c$. The result reads
\begin{equation}
\label{eq:integrate}
\frac{1}{T} \int_{t_c-T/2}^{t_c+T/2} \sqrt{\frac{m}{2 \pi \hbar t}} \exp\left(i \frac{m}{2 \hbar t} (x-x_0-\frac{p}{m} t)^2 - i \frac{\pi}{4}\right) dt = \frac{m}{T p}.
\end{equation}
Note that the final result is the probability density of finding the particle at $x$ under the condition $p$, where the time uncertainty has been limited to a finite value of $T$. It is easy to compare this to a classical ergodic average: within time $T$, the particle will travel a distance of $p T/m$. Therefore, a time uncertainty of $T$ corresponds directly to a position uncertainty of $p T/m$, and the equal distribution over times from $t_c-T/2$ to $t_c+T/2$ results in a probability density of $m/(T p)$ for the position $x$. Thus the ergodic average of the action phase probabilities reproduces the statistics associated with time uncertainties in classical dynamics. 

With regard to energy-time uncertainties, it is important to note that the integral in Eq.(\ref{eq:integrate}) can be approximated by an integral over a complex Gaussian with an imaginary time variance of 
\begin{equation}
\Delta t = \frac{m}{p} \sqrt{\hbar \frac{(x-x_0)}{p}} = \frac{2}{E_p} \sqrt{\hbar p (x-x_0)}.
\end{equation}
The approximation that the particle moves along a geometric line with $x(t)=x_0 + p t/m$ is therefore valid for partial ergodic averages over intervals sufficiently longer than $\Delta t$. Note that this time uncertainty increases with distance, indicating that the relative importance of the initial momentum (and hence the energy) for the final position $x(t)$ increases. In the classical limit, approximate observations of $x_0$ and $x(t)$ provide both energy and time information, where the role of the energy increases with the time difference $t$ between the observations of $x_0$ and $x(t)$. In the microscopic limit of precise temporal resolution, there is no joint reality of time and energy or of position and momentum. At time scales rougher than $\Delta t$, the most precise description is obtained from energy and momentum, while position and time are the relevant concepts at precisions greater than $\Delta t$. This distinction of resolutions explains why elementary particle physics usually ignores the explicit time dependence of particle motion: the experiments in question are all in the regime of negligible time resolution. Unfortunately, such circumstances have shaped the interpretation of quantum mechanics in a way that has seriously damaged our ability to explain the more accessible experience of motion in time. 

In the case of free particles, the classical approximation reproduces the precise results rather well, and there is no qualitative difference between the classical predictions and the correct quantum mechanical expressions. However, the fundamental physics relating energy to time is quite different. Eq.(\ref{eq:integrate}) describes a time integral that involves all propagation times at all energies. It is therefore not possible to argue that a particle needs the energy $E_p$ to get from $x_0$ to $x$ within a specific time $t$. In fact, it is not even possible to argue that the energy $E_p$ must be positive, since Eq.(\ref{eq:integrate}) can be applied to $E_p<0$ as well. The only problem is that $p$ must then be replaced by an imaginary number, which results in a description of tunneling through a barrier of potential energy $V>E$.

For a tunneling barrier of length $L=x-x_0$ and a potential of $V$, the action phase probability of particle motion with $E<V$ is given by
\begin{eqnarray}
\label{eq:tunnelApp}
\lefteqn{P(x(t)|x_0,E) =}
\nonumber \\ &&
 \sqrt{\frac{m}{2 \pi \hbar t}} \exp\left(i \frac{m L^2}{2 \hbar t} - i \frac{V-E}{\hbar} - i \frac{\pi}{4}\right) \exp\left(- \frac{\sqrt{2 m (V-E)}}{\hbar} L\right).
\end{eqnarray}
The action phase of this probability never satisfies the condition given by Eq.(\ref{eq:classE}), since the weak value of energy given by $H(x,x_0,t)$ is always above the barrier. However, there is an analytical solution of the integral, resulting in the ergodic probability of tunneling from $x_0$ to $x$ given by
\begin{eqnarray}
\label{eq:tunnelErg}
\lefteqn{\lim_{T \to \infty} \frac{1}{T} \int P(x(t)|x_0,E) dt =}
\nonumber \\ &&
 \frac{m}{T \sqrt{2 m (V-E)}} \exp\left(- 2 \frac{\sqrt{2 m (V-E)}}{\hbar} L\right).
\end{eqnarray}
Note that the time integral contributes half of the exponential suppression of tunneling, which corresponds to the compensation of the time independent complex phase associated with $p(x-x_0)$ in the ergodic integral for particles moving above the barrier by the complex result of the time integral in Eq.(\ref{eq:integrate}). Quantum ergodicity thus explains tunneling in terms of an energy independent time evolution that is equally valid above and below the barrier. 

It is possible to generalize the concept of tunneling based on the insight that the causality connecting $a$ and $b$ is not limited to contributions with stationary action phases, as suggested by the classical approximation. The idea that a minimal energy $E_n$ is necessary to move from $a$ to $b$ originates from the assignment of an energy to the motion from $a$ to $b$ by $H(b,a,t)=E_n$. If the minimal value of $H(b,a,t)$ for all times $t$ is equal to $V$, then there exists an energy barrier of height $V$ between $a$ and $b$, and no classical trajectory will connect $a$ and $b$ at energies of $E_n<V$.  However, the assignment of a fixed energy to the motion between $a$ and $b$ is merely an approximation used to derive classical trajectories from action phase probabilities. Even for $E_n<V$, the cancellation of complex probabilities caused by the energy mismatch $H(b,a,t)-E_n$ will not be perfect. Therefore, quantum ergodicity predicts a non-vanishing probability of finding $b$ for the time averaged dynamics  starting from $a$ at an energy of $E_n<V$. The logic that prevents tunneling in classical physics merely results in a suppression of the tunneling probability by the time dependence of the action phase probabilities that describe the fundamental causality of motion.

\section{On the interpretation of classical physics}
\label{sec:concepts}

In the course of this work, it has struck me that the reason why we find quantum mechanics so difficult to explain might be that we are reluctant to abandon the idea that differential geometry is the only possible explanation of our intuitive experience of motion \cite{Sch13,Hal14,Bri15}. However, there is no scientific reason to hold on to this idea, since we have absolutely no empirical evidence for the existence of infinitesimal changes in space and time. The truth is that analytical geometry is a mathematical tool that allows us to explain the patterns of causality in our experience, despite the severe limitations in the precision and controllability of that experience. Our assumptions about causality are an essential part of the reconstruction of a reality ``out there'' from what is necessarily incomplete evidence. It is therefore essential that we identify laws of causality that apply equally under all circumstances. 

Does classical physics really need the assumption of trajectories? In the light of the present analysis, it would seem that the answer should be no. The problem is that we may have over interpreted classical physics as a ``theory of reality'' when we should really think of it as a theory of causality. The classical theory of causality emerges naturally as a rough approximation of the more precise description of causality by action phase probabilities. It is only necessary to give up the idea that the motion of an object can be described by an infinitely precise geometrical line of thickness zero. Differential geometry is merely a convenient approximation that applies when the sequences of observation involve only negligible interaction dynamics.

The claim that physical objects are geometric shapes in a spatial and temporal continuum is not supported by any evidence and represents a philosophical statement with dangerously dogmatic implications. Hopefully it is not already too late to cure physics of this dogmatic bias - the history of the interpretation of quantum mechanics is definitely not a success story, and the level of the recent discussions is not very encouraging either. Here, I would like to propose a way out that would address the problem at its very roots. The original quantum revolution started with the observation that the description of motion by differential geometry fails to explain the experimentally observed distribution of energies in physical systems. Specifically, the periodicities of motion appear to correspond to discretization intervals in the energy. It should be emphasized that this is a serious problem for any theory based on differential geometry. In classical trajectories, the value of energy is defined by the instantaneous properties, not by the complete time evolution. Simply by demanding that the energy values depend on the periodicity of the motion, quantum physics provides empirical proof that the mathematics of infinitesimals is only an approximation that breaks down in the limit of $\hbar$. It is therefore not quantum mechanics but classical physics that is in need of a new interpretation. 

The idea that differential geometry describes the motion of physical objects is based on the assumption that physics is scale invariant. This assumption is the basis of all mathematical infinities, and its application to motion resulted in the calculus of infinitesimals that started modern physics. However, the discovery of a fundamental action scale changes the situation. The observation that the values of energy depend on the periodicity of the motion already indicates that the deterministic laws of differential geometry must break down in the limit of small actions. In particular, the assumption that the time evolution of a physical system is a simple analytical function of its energy and other conserved quantities cannot be maintained. Instead, the fundamental relations between energy and time must be expressed by action phase probabilities, where the relation between energy and time is established by the joint action of $E_n t$, while the fundamental time dependence of motion from $a$ to $b$ and the energy dependence of the action phase probability factorize according to Eq.(\ref{eq:parts}). 

The new perspective on the fundamental relation between energy and time that is expressed by action phase probabilities is an important step towards a deeper understanding of the dynamics of physical systems. Historically, it may have been quite damaging to our understanding of quantum mechanics that the original formulation emphasized energy over time and tended to neglect the observation of motion that shapes most of our practical experience of mechanics. Action phase probabilities restore the balance by showing how the effects associated with quantum coherence originate from the fundamental relation between energy and time.

\section{Conclusions}

The main point of the present discussion is the introduction of action phase probabilities as the proper representations of causality in physics. To obtain a consistent explanation of all physical phenomena, we need to replace the classical notion of dynamics as a differential change of physical properties in time with the more appropriate description of causality provided by action phase probabilities. Importantly, this is not a result of purely theoretical reasoning, but a consequence of the evidence obtained from careful experimentation. The assumption that the world can be described by differential geometry broke down when the theory of mathematical trajectories failed to explain energy quantization in physical systems. The problem with the original formulation of quantum theory was that the artificial notion of ``states'' obscured the fact that Hilbert space provides a new and different description of causality. In particular, it is easy to confuse the time evolution of probability given by Schr\"odinger's equation with the deterministic time evolution of classical waves. In fact, the time evolution of physical properties is different from the time evolution of their probabilities, and action phase probabilities are a convenient way of explaining this difference. As shown above, action phase probabilities express exactly the same relations that were previously approximated by classical trajectories, which means that they represent the necessary modifications of classical theory in the light of quantum mechanical evidence. 

With regard to the confusion caused by the role of statistics in quantum physics, it needs to be recognized that classical trajectories are theoretical constructs based on a purely hypothetical availability of complete information about an isolated system. The real purpose of the classical trajectories is that they allow us to judge the causality relations between different parts of the available evidence. In this sense classical determinism never provided us with a microscopic description of reality, we merely tend to over interpret the theory along these lines. According to quantum physics, the need to refer to a complete set of physical properties results in complex valued probabilities, where the probabilities do not describe randomness but universally valid and fully deterministic relations between physical properties. The complex phases of these probabilities represent the dynamics of the system generated by actions that involve the physical properties. In any realistic situation, the attempt to control physical properties results in a trade off described by ergodic averages or partial ergodic averages. Quantum ergodicity thus describes the limits of control and explains why action phase probabilities can provide a fundamental explanation of the deterministic relations between physical properties and of the causality of motion that we observe in the world around us.  


\end{document}